\documentclass[prd,aps,twocolumn,superscriptaddress,preprintnumbers,nofootinbib]{revtex4-1}
\usepackage{pslatex}
\usepackage[pdftex]{graphicx}
\usepackage{psfrag}
\usepackage{epsfig}
\usepackage{color}
\usepackage{cancel}
\usepackage{slashed}
\usepackage{amssymb}
\usepackage{amsmath}
\usepackage{hyperref}
\usepackage{enumerate}
\usepackage{multirow}
\usepackage{natbib}
\usepackage[normalem]{ulem}
\usepackage{mathrsfs}
\DeclareMathAlphabet{\mathpzc}{OT1}{pzc}{m}{it}

\usepackage[english]{babel}

\usepackage{amsmath}
\usepackage{amssymb}

\usepackage{mathrsfs}
\DeclareMathAlphabet{\mathpzc}{OT1}{pzc}{m}{it}

\usepackage{graphicx}

\usepackage{cancel} 
\usepackage{paralist}
\usepackage{booktabs}

\usepackage{hyperref}
\usepackage{xspace}

\usepackage{graphicx,color,verbatim,acronym,epsfig,subfigure}
\usepackage{slashed,skak}
\usepackage{latexsym,amsfonts,amssymb,amsmath,makeidx,mathrsfs,multirow,lineno,bm,bbm}

\usepackage{tikz}
\usepackage{pgfplots}
\usetikzlibrary{trees}
\usetikzlibrary{decorations.pathmorphing}
\usetikzlibrary{decorations.markings}
\usetikzlibrary{decorations.text}
   \definecolor{greeen}{rgb}{0.03,0.54,0.23}
\definecolor{test}{rgb}{0.03,0.74,0.33}
\definecolor{viol}{rgb}{0.44,0,0.94}
\definecolor{or}{rgb}{0.9,0.6,0}
\tikzset{
    photon/.style={decorate, decoration={snake, amplitude=2pt}, draw=green},
    photon2/.style={decorate, decoration={snake, amplitude=2pt}, draw=viol},
    dark/.style={draw=greeen, postaction={decorate},
        decoration={markings,mark=at position .5 with {\arrow[draw=greeen]{>}}}},
antidark/.style={draw=greeen, postaction={decorate},
        decoration={markings,mark=at position .5 with {\arrow[draw=greeen]{<}}}},
electron/.style={draw=viol, postaction={decorate},
        decoration={markings,mark=at position .5 with {\arrow[draw=viol]{>}}}},
        antielectron/.style={draw=viol, postaction={decorate},
        decoration={markings,mark=at position .5 with {\arrow[draw=viol]{<}}}},
        neutrino/.style={draw=orange, postaction={decorate},
        decoration={markings,mark=at position .5 with {\arrow[draw=orange]{>}}}},
        antineutrino/.style={draw=orange, postaction={decorate},
        decoration={markings,mark=at position .5 with {\arrow[draw=orange]{<}}}},
gluon/.style={decorate, draw=or,
        decoration={coil,amplitude=4pt, segment length=4pt}},
  ZZ/.style={decorate, decoration={snake}, draw=yellow},
  scalar1/.style={decorate, dashed, draw=cyan},
  scalar0/.style={decorate, dashed, draw=greeen},
   scalar2/.style={decorate, dashed, draw=viol},
 pseudoscalar/.style={decorate, dashed, draw=purple},
   }

\usetikzlibrary{decorations, decorations.markings, decorations.pathmorphing, arrows, graphs, shapes.geometric, snakes}
\usetikzlibrary{shadings}

\newcommand{\be}{\begin{equation}}
\newcommand{\ee}{\end{equation}}
\newcommand{\bea}{\begin{eqnarray}}
\newcommand{\eea}{\end{eqnarray}}

\newcommand{\abs}[1]{|#1|}

\usepackage{ulem,fancyvrb}
\usepackage{xcolor}

\providecommand{\abs}[1]{\lvert#1\rvert}

\begin{document}
\title{Heavy dark matter and Gravitational waves}
\author{Xin Deng }	
\affiliation{
	~Department of Physics, Chongqing University, Chongqing 401331, China}

\author{Xuewen Liu}
\affiliation{
	~Department of Physics, Yantai University, Yantai 264005, China}

\author{Jing Yang}
\affiliation{
	~Department of Physics, Chongqing University, Chongqing 401331, China}

\author{Ruiyu Zhou }
\affiliation{
	~Department of Physics, Chongqing University, Chongqing 401331, China}

\author{Ligong Bian }
\email{lgbycl@cqu.edu.cn}
\affiliation{
	~Department of Physics, Chongqing University, Chongqing 401331, China}
\begin{abstract}
Domain walls can form after breakdown of a discrete symmetry induced by first-order phase transition, we study the heavy dark matter produced around the temperature of the phase transition that yields the breakdown of a $\mathbb{Z}_{3}$ symmetry. The generated gravitational waves by domain walls decay is found to be able to probed by the Pulsar Timing Arrays, and the future Square Kilometer Array.
\end{abstract}
\maketitle
\section{Introduction}
\label{sec:intro}

The phase transition in the Standard model (SM) is confirmed to be {\it cross-over}~\cite{DOnofrio:2014rug}, and a first-order phase transition (PT) is predicted in many particle physics models that beyond the Standard model (SM)~\cite{Mazumdar:2018dfl}, such as:
 singlet extended SM~\cite{Profumo:2014opa,Zhou:2019uzq,Zhou:2020idp,Alves:2018jsw,Profumo:2007wc,Espinosa:2011ax,Jiang:2015cwa}, two Higgs doublet models~\cite{Cline:2011mm,Dorsch:2013wja,Dorsch:2014qja,Bernon:2017jgv,Andersen:2017ika,Kainulainen:2019kyp}, the George-Macheck model~\cite{Zhou:2018zli}, and the next-to-minimal supersymmetry model~\cite{Bian:2017wfv,Huber:2015znp},etc.
A general prediction of the first-order PT is the production of gravitational waves (GWs)~\cite{Caprini:2015zlo,Caprini:2019egz}, which happens to be
one of the scientific searching goals of the space-based interferometers, including LISA~\cite{Audley:2017drz}, Taiji~\cite{Gong:2014mca}, TianQin~\cite{Luo:2015ght}, DECIGO~\cite{Yagi:2011wg}, and BBO~\cite{Corbin:2005ny}.

Dark matter (DM) is believed to be one crucial ingredient of the Universe, while its nature is still a
mystery. The weak scale WIMP DM currently confront with sever constraints from direct detection DM experiment, such as XENON1T~\cite{Aprile:2018dbl}, PandaX~\cite{Cui:2017nnn}, and LUX~\cite{Akerib:2016vxi}. 
One possibility to alleviate the situation is to consider the
DM production through PT modified thermal freeze-out mechanism as studied in Refs.~\cite{Heurtier:2019beu,Coudarchet:2019auv,Baker:2018vos}, where the dark sector keeps thermal equilibrium with the SM in the early Universe. 
Another scenario is to consider the dark sector feebly interacts with the SM, the relic density can be generated through the freeze-in mechanism~\cite{Hall:2009bx}, where the DM is called FIMPs.
The typical temperature for FIMP DM production is close to the DM mass, in this case the DM production may be highly amplified or diminished by the PT effect, see Refs.~\cite{Baker:2016xzo,Baker:2017zwx,Darme:2020nhh,Bian:2018bxr,Bian:2018mkl}.

 After the PT, the symmetry of the model breaks. One general prediction of the spontaneous breakdown of a discrete symmetry is the formation of the domain walls (DWs)~\cite{Kibble:1976sj}. The collapse of DWs may produce GWs with peak frequency being around nanoHertz~\cite{Kadota:2015dza,Hiramatsu:2013qaa,Bian:2020bps,Zhou:2020ojf}, which can be probed by Pulsar Timing Arrays experiments, such as the European Pulsar Timing Array (EPTA \cite{Desvignes:2016yex}), the Parkes Pulsar Timing Array
(PPTA \cite{Hobbs:2013aka}), the International Pulsar Timing Array (IPTA \cite{Verbiest:2016vem}), and the NANOGrav~\cite{Arzoumanian:2018saf}.  In this paper, we study the fermionic DM production within a $\mathbb{Z}_{3}$ model, which is well-motivated for neutrino and DM physics~\cite{Ma:2007gq,Belanger:2012zr,Cai:2018imb,Arcadi:2017vis,Hektor:2019ote,Kang:2017mkl}. This paper is organized as follows:
We first study the PT in a $\mathbb{Z}_{3}$ model in section~\ref{sec:FTPT}, the section~\ref{secDM} is devoted to the DM study, GWs produced from the first-order PT and DWs decay are studied in section~\ref{sec:gw}, we then summarize in section~\ref{sec:concl}.

\section{ Phase transition model}
\label{sec:FTPT}

In this work, we consider the PT model with the scalar potential given by,
\begin{align}
  V = \mu_{S}^{2} \abs{S}^{2} + \lambda_{S} \abs{S}^{4} + \frac{\mu_3}{2} (S^{3} + S^{\dagger 3})\;,
\end{align}
where $S=(  v_s + s + i A)/\sqrt{2}$, and $v_s$ is the VEV of $s$ field.
The cubic term  $\mu_{3}$ breaks the global $U(1)$ symmetry (under transformation $S \to e^{i \alpha} S$ ) with
a remanent unbroken $\mathbb{Z}_{3}$ symmetry. Considering the stationary point condition,
\begin{align}
\left.\frac{d V(s, A)}{d s}\right|_{s=v_s}=0\;,
\end{align}
we get $\mu_s^2=-\lambda_S v_s^2  - \frac{3\sqrt{2}}{4} \mu_3 v_s $.
The masses of the new particles (s, A) are then given by,
\begin{align}
m_s^2&=\mu_s^2 + 3\lambda_s v_s^2+\frac{3\mu_3 v_s}{\sqrt{2}} \;,\nonumber \\
m_A^2 &= \mu_s^2 + \lambda_s v_s^2-\frac{3\mu_3 v_s}{\sqrt{2}} \;.\nonumber
\label{eq:mass:matrix}
\end{align}
The cubic interaction coupling $\mu_3$ and the VEV of $s$ relate to the physical masses through,
\begin{align}
  v_s = \sqrt{\frac{3 m_s^2+m_A^2}{6\lambda_s}} \;,
  \mu_3 = -\frac{4m_A^2\sqrt{2\lambda_s(3 m_s^2+m_A^2)}}{9\sqrt{6}m_s^2+3\sqrt{6} m_A^2}\;.
\end{align}


The finite-temperature effective potential at one-loop level is given by
\begin{align}
  V_\text{eff}(s,T) = V_\text{0}(s)
                      + V_\text{CW}(s)+ V_\text{ct}(s)
                      + V_T(s,T)  \,,
\end{align}
 Here, $V_\text{0}(s)$ is the tree-level potential in terms of the classical field,
 \begin{align}
 V_\text{0}(s)=\frac{\mu_s^{2}}{2}s^{2}+\frac{\mu_3}{2 \sqrt{2}}s^{3}+\frac{\lambda_{S}}{4}s^{4}\;.
 \end{align}
 The one-loop
contribution splits up into a ultraviolet-divergent zero-temperature
(Coleman--Weinberg)
part, $V_\text{CW}$, and a finite-temperature part, $V_T$.  We also include the
resummed contribution from ring diagrams, $V_\text{daisy}$.

The one-loop Coleman-Weinberg potential is~\cite{Coleman:1973jx,
Quiros:1999jp}
\begin{align}
  V_\text{CW}(s) &= \sum_i \frac{\eta_i n_i}{64 \pi^2} m_i^4(s)\bigg[\log\bigg(\frac{m_i^2(s)}{\Lambda^2}\bigg) - C_i\bigg]\;,
\end{align}
where $i$ runs over $s$-dependent mass of all particle species, $n_i$ is the
number of degrees of freedom for each species,
$\eta_i = +1$ ($-1$) for scalars (fermions), and $C_i = 3/2$ ($5/6$) for scalars and fermions.
Here, we take $\Lambda=v_s$.
The counter-terms are introduced to prevent the shift of the vacuum driven by the Coleman-Weinberg potential,
with
\begin{align}
  V_\text{ct}(s) &=  \frac{\delta\mu_s^2}{2} s^2
+ \frac{\delta\mu_3}{2\sqrt{2}} s^3
+ \frac{\delta\lambda_s}{4} s^4\;, 
\end{align}
where, these terms are fixed through following conditions,
\begin{align}
& \frac{d (V_\text{CW}(s)+ V_\text{ct}(s))}{d s}\bigg|_{s=v_s} = 0 \,,
 \frac{d^2 (V_\text{CW}(s)+ V_\text{ct}(s)) }{d s^2}\bigg|_{s=v_s}= 0 \,, \nonumber\\
& V_\text{CW}(0) - V_\text{CW}(v_s)- V_\text{ct}(v_s)= 0 \,.
\end{align}

The one-loop finite-temperature correction is evaluated
to be~\cite{Dolan:1973qd, Quiros:1999jp}
\begin{align}
  V_T(s,T) &=\frac{T^4}{2\pi} \sum_i n_i J_{B,F}\bigg( \frac{m_i^2(s,T)}{T^2}\bigg)\,,
\end{align}
where the function $J_{B,F}$ is given by
\be
\label{eq:jfunc}
J_{B,F}(y) = \pm \int_0^\infty\, dx\, x^2\, \ln\left[1\mp {\rm exp}\left(-\sqrt{x^2+y}\right)\right]\; ,
\ee
with the upper (lower) sign corresponding to bosonic (fermionic) contributions.
Here, the thermal masses of the scalars are
\begin{align}\label{eqms}
  m_s^2(s,T) &= m_s^2+\frac{4\lambda_S}{12} T^2 \,, \\
  m_A^2(s,T) &= m_A^2+\frac{\lambda_S}{3} T^2 \,.
\end{align}

The onset condition of the first-order phase transition is ~\cite{Affleck:1980ac,Linde:1981zj,Linde:1980tt}:
\begin{eqnarray}\label{eq:bn}
\Gamma\approx A(T)e^{-S_3(T)/T}\sim 1\;,
\end{eqnarray}
which means the nucleation temperature ($T_n$) is obtained when the number of bubbles for bubble nucleation per horizon volume and per horizon time is of order unity. Here,
\begin{eqnarray}
S_3(T)=\int 4\pi r^2d r\bigg[\frac{1}{2}\big(\frac{d s}{dr}\big)^2+V_{\rm eff}(s,T)\bigg]\;,
\end{eqnarray}
can be extremized to find the solution of the bounce of configuration of the nucleated bubbles, i.e., the bounce configuration of the field connects the $\mathbb{Z}_3$ broken vacuum (true vacuum) and the $\mathbb{Z}_3$ preserving vacuum (false vacuum), with the boundary conditions
\begin{eqnarray}
\lim_{r\rightarrow \infty}s  =0\;, \quad \quad \frac{ds }{d r}|_{r=0}=0\;.
\end{eqnarray}

There are two crucial parameters for GWs calculation: 1) the latent heat of the first-order PT normalized by the radiative energy, i.e., $\alpha=\frac{\Delta\rho}{\rho_R}\;$ with
$\Delta \rho$ being the released latent heat from the phase transition to the energy density of the plasma background; 2)
the inverse time duration of the phase transition, which is defined as
\begin{eqnarray}
\frac{\beta}{H_n}=T\frac{d (S_3(T)/T)}{d T}|_{T=T_n}\; .
\end{eqnarray}
We show three benchmarks estimated by analyzing the PT dynamics utilizing {\tt CosmoTransition}~\cite{Wainwright:2011kj}
 in table~\ref{tabp2}, which will be used for DM and GWs study in the following sections.

\begin{table}[!htp]
\footnotesize
\begin{tabular}{c c c c c c c c c }
\hline
BMs~&~$\lambda_s$~&~~$m_s~(GeV)$~ &~$m_A$~(GeV)~ & ~$v_s(T_n)$~(GeV)~ & ~ $T_n$~(GeV)~ & ~ $\alpha$~ & ~ $\beta/H_n$  \\
\hline
$BM_1$  & $1.52$ & $14991.4$ & $40278.8$ & $14878.4$ & $12235.2$ & $0.005$ & $14310.7$ \\

$BM_2$  & $1.27$ &$7889.7$&$21655.6$&$8859.8$&$6291.1$& $0.006$ & $6991.3$ \\

$BM_3$  & $0.35$ &$3843.2$&$9959.3$&$7686.8$&$4856.6$& $0.004$ & $5428.6$ \\

\hline
\end{tabular}
\caption{The benchmark points for achieving the first-order PT.}
\label{tabp2}
\end{table}

\section{Dark matter }
\label{secDM}

Refs.~\cite{Chiang:2019oms,Kannike:2019mzk} considered the weak scale WIMP DM being the pseudoscalar $A$ and analyzed the GWs production at first-order electroweak PT. 
In this paper, we are going to study fermionic DM production at the high scale PT assisted by the $A$.
Concretely, we consider two Dirac fermions interacting with the $\mathbb{Z}_3$ scalar through the Lagrangian,
\begin{align}
   \mathscr{L} \supset m_{\psi_1} \bar\psi_1\psi_1+m_{\psi_2} \bar\psi_2 \psi_2+ y_{s} S \psi_1 \bar \psi_2 +y_{s} S^{\dagger} \bar\psi_1 \psi_2\;.
\end{align}
The complex scalar and the two fermionic particles have $\mathbb{Z}_3$ charges: 1,1 -1.
After the $\mathbb{Z}_3$ symmetry breakdown, the two Dirac fermions $\psi_{1}$ and $\psi_{2}$ mixed each other and the lightest one may serve as DM candidate (denoted as $\chi$),
the mass matrix is given by,
\begin{align}
\left(
\begin{array}{cc}
 \frac{1}{2} \left(-\sqrt{M}+m_{\text{$\psi_1 $}}+m_{\text{$\psi_2 $}}\right) & 0 \\
 0 & \frac{1}{2} \left(\sqrt{M}+m_{\text{$\psi_1 $}}+m_{\text{$\psi_2 $}}\right) \\
\end{array}
\right)
\end{align}
where $M=-2 m_{\text{$\psi_1$}} m_{\text{$\psi_2 $}}+m_{\text{$\psi_1 $}}^2+m_{\text{$\psi_2$}}^2+4 y_s^2 v_s^2(T)$.

\begin{figure}[!htp]
\begin{center}
\includegraphics[width=0.43\textwidth]{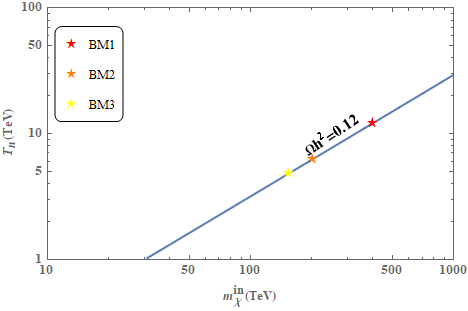}
\caption{The solid line is calculated by Eq.~\ref{eq:Omega_chi} for $\Omega h^2=0.12$. The stars  correspond to the benchmark points in Table~\ref{tabp2}.}
\label{Omega}
\end{center}
\end{figure}

We first consider the case where fermionic DM particles
$\chi$ can freeze out, while the dark scalars remain
in thermal equilibrium with the SM bath. As in Refs.~\cite{Baker:2019ndr,Chway:2019kft,Marfatia:2020bcs,Hong:2020est}, we assume the DM keep in thermal equilibrium with the thermal bath, being massless outside the bubble and massive inside the bubble.  As DM particles enter the bubble, their
interactions are put abruptly out of equilibrium, that prevent their
annihilation and therefore constitute the DM.
Considering $m_\chi = x_f T_n\sim 20 T_n$,  one has the DM mass $m_\chi\sim10^{4-6}$ GeV for the nucleation temperature $T_n\sim10^{3-5}$ GeV.
With the non-equilibrium explicitly given by the condition,
\begin{align}
  \frac{m_\chi^\mathrm{in}}{T_n}
    &\gtrsim 24 - \log \frac{T_n}{\text{TeV}} - \frac{3}{2} \log\frac{m_\chi^\mathrm{in}/T_n}{24}
                + 4 \log y_s \,,
  \label{eq:OOE-condition}
\end{align}
and the DM relic density calculated as
\begin{align}
  \Omega_\mathrm{DM} h^2
    &\simeq
    0.17 \,
    \bigg( \frac{T_n}{\text{TeV}} \bigg)
    \bigg( \frac{m_\chi^\mathrm{in} / T_n}{30} \bigg)^{5/2}
    \frac{e^{- m_\chi^\mathrm{in} / T_n}}{e^{-30}} \,,
  \label{eq:Omega_chi}
\end{align}
where the $m_\chi^\mathrm{in}=y_s v_s(T_n)$ with $m_{\text{$\psi_1$}}=m_{\text{$\psi_1 $}}=0$.
The observed relic abundance is obtained for $m_\chi$ and $T_n$ corresponding to the solid line in Fig.~\ref{Omega}. For the PT temperature shown in the three benchmark points in Fig.~\ref{Omega} (red, orange, yellow stars),  we found the Yukawa couplings are required by the out-of-equilibrium condition (Eq.~\ref{eq:OOE-condition}) to be: $y_s=20.5,14.4, 12.5$, which is too large and will break the perturbativity.   

To address this contradiction, we then consider a feebly interaction strength of the Yukawa $y_s$,
and the complex scalar $S$ keep in thermal equilibrium with the SM, we therefore can consider the DM particle $\chi$ as a FIMP.
DM may be produced through the decay channel $s\rightarrow\bar{ \psi}\chi$, and the scattering channels $s(A) s(A)\rightarrow\bar{\psi} \chi$ via s-channel with mediator being $s$, and the t-channel of $ss\rightarrow \chi\bar{\chi}$ as shown
in the Fig.~\ref{Feynmanpicture}. The annihilation cross sections and decay width for these processes are given in the Appendix~\ref{sec:appendix}. Introducing the dimensionless evolution variable $x\equiv m_\chi /T$, the Boltzmann equation that governs the evolution of DM number density (here, $Y_i \equiv n_i/s$) is given by\;,
\begin{eqnarray}
\frac{{dY}_{\chi }}{{dx}}&=&\frac{m_{{s}}^2 g_s}{2 \pi ^2}\frac{m_{\chi }}{H(x)s(x)x^2}\Gamma(s\rightarrow \chi \overset{-}{\psi })K_1(\frac{x m_{\text{s}}}{m_\chi })+\frac{g_{B_1}^2}{32 \pi ^4}\frac{m_{\chi }}{H(x)s(x)x^2}\;\nonumber\\
&\times&\int_{4 m_{B_1}}^{\infty} d \mathfrak{s}(\mathfrak{s}-4 m_{B_1}^2)\sigma(B_1\overset{-}{B}_1\to \text{$\chi $B}_2) \sqrt{\mathfrak{s}}K_1(\frac{x \sqrt{\mathfrak{s}}}{m_{\chi}})\;,
\end{eqnarray}
where $H(x)$ denotes the Hubble rate, $\sqrt{\mathfrak{s}}$ is the centre-of-mass energy, $g_s$($g_{B_1}$) is the number of degrees of freedom of s($B_1$) and $K_1$ is a modified Bessel function of second kind. As in Refs.~\cite{Baker:2016xzo,Baker:2017zwx,Darme:2020nhh,Bian:2018bxr,Bian:2018mkl}, to capture the PT effects, we calculate the DM number density by the replacement of $m_{s,Z}\to m_{s,A}(T)$ with the thermal masses of $m_{s,A}(T)$ being given by Eq.~\ref{eqms}.

\begin{figure}[!htp]
\begin{center}
\includegraphics[width=0.5\textwidth]{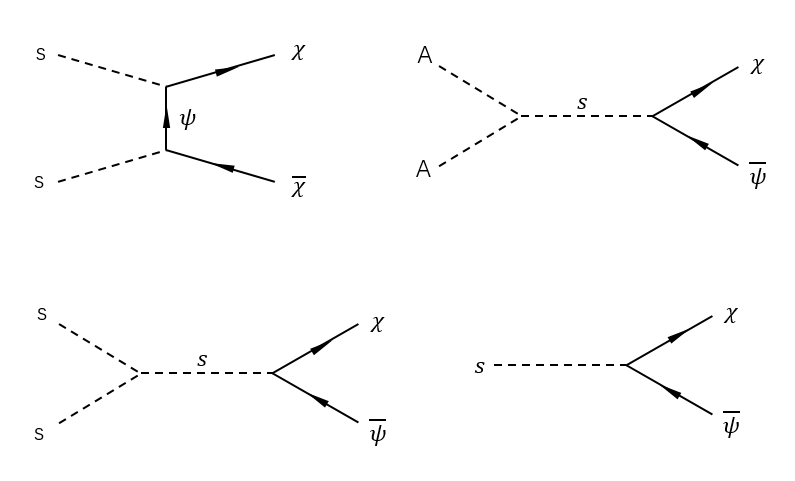}
\caption{Processes relevant for the freeze-in mechanism of the DM. }\label{Feynmanpicture}
\end{center}
\end{figure}

\begin{figure}[!htp]
\begin{center}
\includegraphics[width=0.4\textwidth]{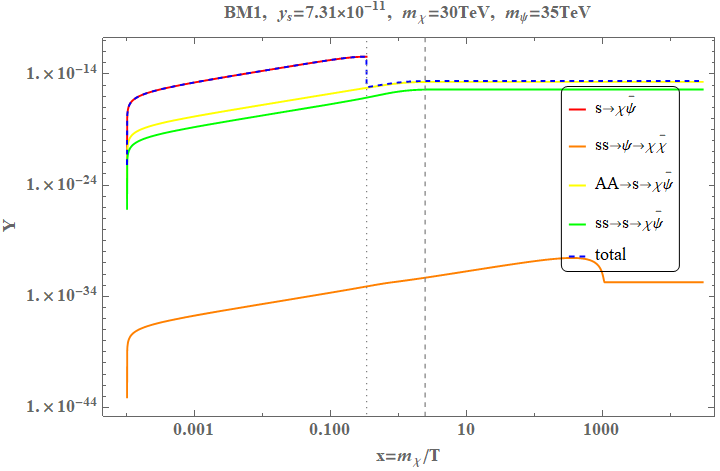}
\includegraphics[width=0.4\textwidth]{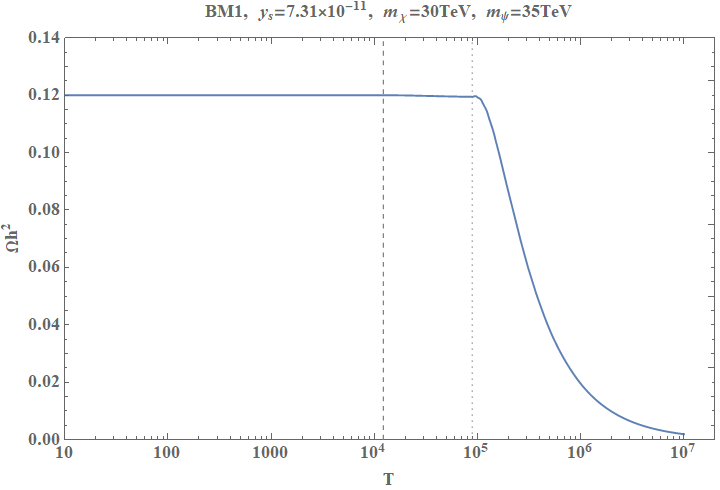}
\caption{Top: The DM number density evolution of different channels. Bottom: The DM relic density $\Omega h^2$ as a function of temperature T. The vertical dotted and dashed line correspond to $m_s(T)=m_ \psi + m_ \chi$ and $T_n$.}\label{relic}
\end{center}
\end{figure}

To illustrate the DM production process at the first-order PT that induces DW formation, we calculate the DM relic density evolution in Fig.~\ref{relic} by taking input parameters from the BM$_1$ in Table.~\ref{tabp2}. 
As shown in the top panel of Fig.~\ref{relic}, there were four channels can contribute to DM number density. The mass $m_s(T)$ decreases with temperature, as indicated by Eq.~\ref{eqms}, the decay channel becomes active and dominates the DM production process before the PT ($T>T_n$). As the Universe cools down, bubble nucleate, the $\mathbb{Z}_3$ symmetry break spontaneously and the DWs form. Just below the nucleation temperature (T=8.9 TeV),  the DM production from the decay channel stops because of $m_s(T)<m_ \psi + m_ \chi$, then the scattering channels dominate the DM production. The bottom panel of Fig.~\ref{relic} shows that the DM relic density is mostly produced by the decay channel, and stop increase when $T< 8.9$ TeV after the PT. 

\section{Stochastic GWs}
\label{sec:gw}

In this section, we calculate stochastic GWs from first-order PT and DWs decay produced after the spontaneous symmetry breaking of the $\mathbb{Z}_3$ symmetry at high scales. DM relic density accumulates from freeze-in mechanism affected by PT in the same stage. 

\subsection{ GWs from first-order PT}

For the GWs produced from the first-order PT, we focus on the dominate source, i.e., sound waves created in the plasma.
The energy density spectrum is given by~\cite{Caprini:2015zlo}
\begin{align}
\Omega h^2_{\rm sw}(f)&=2.65 \times 10^{-6}(H_*\tau_{sw})\left(\frac{\beta}{H}\right)^{-1} v_b
\left(\frac{\kappa_\nu \alpha }{1+\alpha }\right)^2\nonumber\\
&\times \left(\frac{g_*}{100}\right)^{-\frac{1}{3}}
\left(\frac{f}{f_{\rm sw}}\right)^3 \left(\frac{7}{4+3 \left(f/f_{\rm sw}\right)^2}\right)^{7/2}\;.
\end{align}
We include $\tau_{sw}=min\left[\frac{1}{H_*},\frac{R_*}{\bar{U}_f}\right]$ with $H_*R_*=v_b(8\pi)^{1/3}(\beta/H)^{-1}$ to consider the duration of the phase transition~\cite{Ellis:2020awk}. The root-mean-square (RMS) fluid velocity is estimated as \cite{Hindmarsh:2017gnf, Caprini:2019egz, Ellis:2019oqb}
\begin{equation}
\bar{U}_f^2\approx\frac{3}{4}\frac{\kappa_\nu\alpha}{1+\alpha}\;.
\end{equation}
The efficiency factor $\kappa_\nu$ describes the latent heat transferred into the kinetic energy~\cite{Espinosa:2010hh}. We use bubble wall velocity $v_b=1$ for this study.\footnote{We note that this velocity is relatively larger that that of the EWBG calculations~\cite{Zhou:2020idp,Zhou:2019uzq,Zhou:2020xqi,Alves:2018oct,Alves:2018jsw,Alves:2019igs}. }

\subsection{GWs from DWs decay}

For the DWs decay, we consider the gravitational radiation produced in the radiation dominated era.
The spectrum of the GWs for the DWs decay is estimated to be $\Omega_{GW}^{dw}h^2 \propto f^3$ when $f \textless f_{peak}$, and $\Omega^{dw}_{GW}h^2 \propto f^{-1}$ when $f \geqslant f_{peak}$~\cite{Hiramatsu:2013qaa}.
The amplitude of the generate GWs at the present time $t_0$ is~\cite{Hiramatsu:2013qaa,Kadota:2015dza}
\begin{align}
\Omega^{dw}_{\mathrm{GW}} h^{2}\left(t_{0}\right)_{\mathrm{peak}} & \simeq 5.20 \times 10^{-20} \times \tilde{\epsilon}_{\mathrm{gw}} \mathcal{A}^{4}\left(\frac{10.75}{g_{*}}\right)^{1 / 3}\nonumber\\
& \times \left(\frac{\sigma_{\mathrm{wall}}}{1 \mathrm{TeV}^{3}}\right)^{4}\left(\frac{1 \mathrm{MeV}^{4}}{\Delta V}\right)^{2}.\label{eq:gwdw}
\end{align}
at the peak frequency, which is given by the Hubble parameter at the decay time~\cite{Hiramatsu:2013qaa}:
\begin{align}
f^{dw}&\left(t_{0}\right)_{\mathrm{peak}}=\frac{a\left(t_{\mathrm{dec}}\right)}{a\left(t_{0}\right)} H\left(t_{\mathrm{dec}}\right) \nonumber\\
& \simeq 3.99 \times 10^{-9} \mathrm{Hz} \mathcal{A}^{-1 / 2}\left(\frac{1 \mathrm{TeV}^{3}}{\sigma_{\mathrm{wall}}}\right)^{1 / 2}\left(\frac{\Delta V}{1 \mathrm{MeV}^{4}}\right)^{1 / 2}\;,\label{eq:gwfp}
\end{align}
where the area parameter $\mathcal{A}=1.2$ for the $\mathbb{Z}_3$ symmetry potential~\cite{Kadota:2015dza}. We take the efficiency parameter $\tilde{\epsilon}_{\mathrm{gw}}=0.7$~\cite{Hiramatsu:2013qaa}, and the degree of freedom $g_{*}= 10.75$ at the decay time of the domain walls~\cite{Kadota:2015dza}. The bias term $\Delta V$ explicitly breaks the $\mathbb{Z}_3$ symmetry and therfore determines the decay time of the DW,
\begin{equation}
t_{dec}\approx \mathcal{A}\sigma_{wall}/(\Delta V)\;.
\end{equation}
To avoid overclosing Universe, the domain wall decay should fulfill the following relation,
\begin{eqnarray}
\sigma_{wall}< 2.93\times 10^4 \mathrm{TeV}^3 \mathcal{A}^{-1} (0.1 sec/t_{dec})\;.
\end{eqnarray}

Further, the lower limit on the magnitude of the bias term can be obtained by requiring the DWs decay before the BBN with $t_{dec}\leq 0.01 sec$~\cite{Kawasaki:2004yh,Kawasaki:2004qu},
\begin{eqnarray}
 \Delta V \gtrsim 6.6 \times 10^{-2}\mathrm{MeV}^{4} \mathcal{A} \left(\frac{\sigma_{wall}}{1 \mathrm{TeV}^{3}}\right)\;.
\end{eqnarray}
Here, the magnitude of the bias term and that of the potential around the core of DWs should satisfy the condition of $\Delta V \ll V$, in order not to affect the PT dynamics. The DW solution~\cite{Vilenkin:2000jqa} and the calculation of the DW tension are as follows.
Introducing the phase of the singlet as $S=v_s e^{i\phi}$, we obtain the potential of $\phi$,
\begin{align}
  V &=
  \frac{\mu_{S}^{2}}{2} v_s^{2} + \frac{\lambda_{S}}{4} v_s^{4} + \frac{\mu_3}{2\sqrt{2}} v_s^3 \cos(3\phi) \;.
\end{align}
Then kinetic term of $\phi$ is,
\begin{eqnarray}
\mathcal{L}_{\text {kinetic }}(\phi)=\eta^{2}\left(\partial_{\mu} \phi\right)\left(\partial^{\mu} \phi\right)\;,
\end{eqnarray}
where, $\eta^2=v_s^2/2$.
The field equation of
\begin{eqnarray}
\partial_{\mu} \frac{\partial \mathcal{L}_{\text {kinetic }}}{\partial_{\mu}(\partial \phi)}+\frac{\partial V}{\partial \phi}=0\;
\end{eqnarray}
yields
\begin{eqnarray}
\frac{\mathrm{d}^{2} \phi}{\mathrm{d} z^{2}}-\frac{1}{3 B^{2}} \sin (3 \phi)=0\;,
\end{eqnarray}
where, we note
\begin{eqnarray}
\frac{1}{B^{2}}=-\frac{9}{4}\mu_3 v_s^2\;, \phi = \frac{4}{3} \arctan(e^{\frac{z}{B}})\;.
\end{eqnarray}
We then consider a planar DW orthogonal to the z-axis~\cite{Hattori:2015xla}, i.e., $\phi(z)$, and estimate the DW tension as,
\begin{align}
\sigma_{wall}&=\int d z \rho_{\text {wall }}(z)& \nonumber\\
&=\int\bigg(\bigg|\frac{dS}{dz}\bigg|^2+V\bigg(\frac{S(z)}{\sqrt{2}}\bigg)-V\bigg(\frac{v_s}{\sqrt{2}}\bigg)\bigg) d z\;.
\end{align}

\subsection{Numerical results}

\begin{figure}[!htp]
\begin{center}
\includegraphics[width=0.4\textwidth]{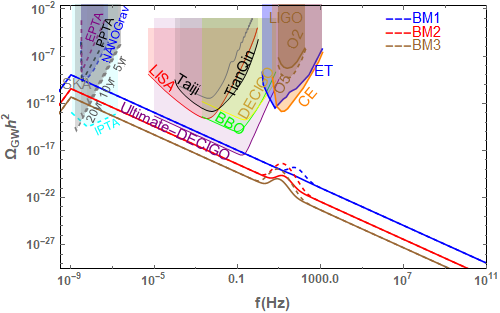}
\caption{Sensitivities of GW detectors on GWs from the first-order PT and the DWs decay. }\label{DWGW-singal}
\end{center}
\end{figure}

We explore the GWs signal from DWs decay after the PT at high scale where the $\mathbb{Z}_3$ symmetry break spontaneously.
Fig.~\ref{DWGW-singal} depicts that, for the benchmarks under study, the peak frequency of GWs from DW decay locates around nanoHertz and can be probed by future PTA and Square Kilometer Array experiments. Due to large plasma energy at high PT temperatures, amplitude of GWs from the PT for these benchmarks under study (see Table~\ref{tabp2}) is too low to be reached by future space-base interferometers, such as: LIGO~\cite{Abbott:2016blz,Aasi:2014mqd,Thrane:2013oya,LIGOScientific:2019vic}, Einstein Telescope \cite{Hild:2010id, Punturo:2010zz}, and Cosmic Explorer~\cite{Evans:2016mbw}. The data of NANOGrav 11yr restrict the phase transition temperature to be $T_n\leq\mathcal{O}(10^4)$ GeV. For the constraint on DW from the NANOGrav 12.5yr data, we refer to Ref.~\cite{Bian:2020bps}.

\section{Conclusion and discussions}
\label{sec:concl}

In this work, we study the DM production during the first-order PT 
and explore GWs from DWs decay formed after the spontaneously symmetry breaking of a $\mathbb{Z}_3$ symmetry that is induced by first-order PT. 
Our study shows that: 
1) fermionic DM mass can be generated though the first-order PT, we found the so-called filtered DM mechanism cannot be realized during the PT process due to the large Yukawa couplings, which are required to fulfill the decouple condition; 
2) heavy fermionic DM ($m_\chi\sim \mathcal{O}(10)$ TeV) can be produced around the temperature of the PT, where the particle mass threshold and DM production channels are altered via the temperature dependent masses.
The GW density can be a two-peak spectrum, the PTA experiments with the peak frequency sensitivity around nanoHertz can be used to probe the peak frequency that corresponds to the symmetry breaking scale of the discrete symmetry.

\section{Acknowledgements}

We appreciate Joachim Kopp, Jim Cline, Benoit Laurent, Danny Marfatia, and Ke-Pan Xie for helpful communications. 
Ligong Bian was supported by the National Natural Science Foundation of China under the grants Nos.12075041, 11605016, and 11947406, and Chongqing Natural Science Foundation (Grants No.cstc2020jcyj-msxmX0814), and the Fundamental Research Funds for the Central Universities of China (No. 2019CDXYWL0029). Xuewen Liu was supported by the National Natural Science Foundation of China under the grants Nos.11947034 and 12005180, and by the Natural Science Foundation of Shandong Province under the grants No. ZR2020QA083.

\section{Appendix}
\label{sec:appendix}

The decay width and annihilation cross sections for the processes shown in Fig.~\ref{Feynmanpicture} are:
\begin{align}
\Gamma(s\rightarrow \chi \overset{-}{\psi })&=\frac{y_{s}^2}{8 \pi }\frac{m_s^2-\left(m_{\chi }+m_{\psi }\right){}^2}{m_s^3} \nonumber\\
& \times \sqrt{\left[m_s^2-\left(m_{\chi }+m_{\psi }\right){}^2\right]\left[m_s^2-\left(m_{\chi }-m_{\psi }\right){}^2\right]}\;,
\end{align}
\begin{align}
\sigma(\text{AA}\to &s\to \chi \overset{-}{\psi })=\frac{9 \mu_3^2 y_{s}^2 \left(E_{\text{cm}}{}^2-\left(m_{\chi }+m_{\psi }\right){}^2\right) }{64 \pi  E_{\text{cm}}{}^3 \sqrt{E_{\text{cm}}{}^2-4 m_A^2} \left(E_{\text{cm}}{}^2-m_s^2\right){}^2} \nonumber\\
&\times \sqrt{-2 E_{\text{cm}}{}^2 \left(m_{\chi }^2+m_{\psi }^2\right)+\left(m_{\chi }^2-m_{\psi }^2\right){}^2+E_{\text{cm}}{}^4}\;,
\end{align}
\begin{align}
\sigma(\text{ss}\to &s\to \chi \overset{-}{\psi })=\frac{\mu _3^2 y_{s}^2 \left(E_{\text{cm}}{}^2-\left(m_{\chi }+m_{\psi }\right){}^2\right)}{32 \pi  E_{\text{cm}}{}^3 \sqrt{E_{\text{cm}}{}^2-4 m_s^2} \left(E_{\text{cm}}{}^2-m_s^2\right){}^2} \nonumber\\
&\times \sqrt{-2 E_{\text{cm}}{}^2 \left(m_{\chi }^2+m_{\psi }^2\right)+\left(m_{\chi }^2-m_{\psi }^2\right){}^2+E_{\text{cm}}{}^4}\;,
\end{align}
\begin{widetext}
\begin{align}
\sigma(\text{ss}\to \psi \to \chi \overset{-}{\chi })&=\frac{y_{s}^4}{8 \pi  E_{\text{cm}}{}^2} \times \left(-\left(\left(2\sqrt{E_{\text{cm}}{}^2-4 m_{\chi }^2}\left(4 m_{\chi } m_{\psi } \left(m_{\chi }-m_s\right) \left(m_s+m_{\chi }\right)+3 (m_s^2-m_{\chi }^2)^2 \right.\right.\right.\right. +4 m_{\chi } m_{\psi }^3+3 m_{\psi }^4 \nonumber\\
&\left.\left.\left.\left.\left.+2 m_{\psi }^2 \left(E_{cm}{}^2-3 m_s^2+m_{\chi }^2\right)\right)\right)\right/ \left(\sqrt{E_{cm}{}^4-4 m_s^2} \left(m_{\psi }^2 \left(E_{cm}{}^2-2 \left(m_s^2+m_{\chi }^2\right)\right)+\left(m_s^2-m_{\chi }^2\right){}^2+m_{\psi }^4\right)\right)\right)\right. \nonumber\\
&-\left(\left(\log \left(-\sqrt{E_{\text{cm}}{}^2-4 m_s^2} \sqrt{E_{\text{cm}}{}^2-4 m_{\chi }^2}-2 \left(m_s^2+m_{\chi }^2-m_{\psi }^2\right)+E_{\text{cm}}{}^2\right)\right.\right.\nonumber\\
&-\log \left(\sqrt{E_{\text{cm}}{}^2-4 m_s^2} \sqrt{E_{\text{cm}}{}^2-4 m_{\chi }^2}-2 \left(m_s^2+m_{\chi }^2-m_{\psi }^2\right)+E_{\text{cm}}{}^2\right)\nonumber\\
&-\log \left(-\sqrt{E_{\text{cm}}{}^2-4 m_s^2} \sqrt{E_{\text{cm}}{}^2-4 m_{\chi }^2}+2 \left(m_s^2+m_{\chi }^2-m_{\psi }^2\right)-E_{\text{cm}}{}^2\right)\nonumber\\
&\left.+\log \left(\sqrt{E_{\text{cm}}{}^2-4 m_s^2} \sqrt{E_{\text{cm}}{}^2-4 m_{\chi }^2}+2 \left(m_s^2+m_{\chi }^2-m_{\psi }^2\right)-E_{\text{cm}}{}^2\right)\right)\nonumber\\
&\times \left(E_{cm}{}^4+2 \left(m_s-m_{\chi }-m_{\psi }\right) \left(m_s+m_{\chi }+m_{\psi }\right) \left(3 m_s^2+\left(m_{\chi }+m_{\psi }\right) \left(5 m_{\chi }-3 m_{\psi }\right)\right)\right.\nonumber\\
&\left.\left.\left.\left.\left.+E_{cm}{}^2 \left(8 m_{\psi } \left(m_{\chi }+m_{\psi }\right)-4 m_s^2\right)\right)\right)\right)/\left((E_{cm}{}^4-4 m_s^2\right) \left(E_{cm}{}^4-2 \left(m_s^2+m_{\chi }^2-m_{\psi }^2\right)\right)\right)\right)\;.
\end{align}
\end{widetext}

\end{document}